# Non-homogeneities in the spatial distribution of gamma-ray bursts

A.A. Raikov
Main (Pulkovo) Astronomical Observatory, St. Petersburg, Russia

V.V. Orlov, O.B. Beketov
St. Petersburg State University, St. Petersburg, Russia

In order to reveal possible non-homogeneities in the spatial distribution of long ($T_{90} > 2^s$) gamma-ray bursts we have examined 201 of them with known redshifts. For different functional forms of metric distance $r(z)$, we use the distribution function $f(l)$ of separations between objects. Asymptotically, for small $l$ this function behaves like $f(l) \sim l^{D-1}$ for fractal sets whose fractal dimension is $D$. It is revealed that for all of the considered forms of $r(z)$ the spatial distribution of gamma-ray burst sources shows signs of fractality with $D = 2.2 \div 2.5$. A few spatially isolated groups of gamma-ray burst sources have been found, one of which has equatorial coordinates $\alpha$ extending from $23^h 56^m$ to $0^h 49^m$ and $\delta$ from $+19^o$ to $+23^o$; its redshifts being within the range of 0.81 to 0.94.

## INTRODUCTION

To study large-scale structure of the Universe, one uses statistical analysis of samples of various radiation sources. Those are the galaxies, galaxy clusters, quasars, and supernovae. Depending on a chosen type of radiation sources, one can investigate the Universe structure on different scales. For example, using the galaxies one may follow three-dimensional structures at $z < 0.2$, however at larger $z$ the incompleteness of catalogues takes place. Other objects (e.g., quasars and supernovae) are not suitable for search and study of large-scale space structures because of strong non-homogeneity of the samples. However, note that the supernovae are the most precise indicators of photometric distances ("standard candles").

In this paper, we firstly purpose to study the large-scale spatial distribution of matter using the gamma-ray burst sources with measured $z$ as "markers". A number of registered gamma-ray bursts is $\sim 10^4$ events, the coordinates on celestial sphere have been measured for more than 1000 objects, and the redshifts are known for 230 objects (at December 10, 2009). It is evident that such sample is rather small to use the standard methods for analysis of large-scale strucutures (see, e.g., Gabrielli et al. 2005, Martinez & Saar 2002, Baryshev & Teerikorpi 2005, Vasilyev 2008), among which are the following:
1) that of conditional density – the counts of objects within the spheres of different radii which centers are placed in the sample objects;
2) that of two-point correlation functions;
3) that of nearest neighbor using the mean distance to it.

In this paper, we suggest a new method based on analysis of distribution of *paired separations*, that allows to derive the quantitative estimations of fractal dimension $D$ using even rather small samples, such as the sample of gamma-ray burst sources with known $z$.

## METHOD OF PAIRED SEPARATIONS

The approach consists in calculation of distribution $f(l)$ of paired separations for the sets of points distributed randomly within the hypersphere of integer dimension (the details are discussed in the book by Kendall & Moran 1963, see also references therein). In particular, it has been shown that the distribution $f(l)$ for the hypersphere of dimension $D$ has the form:

$$f(l) = D l^{D-1} (L/2)^{-D} I_\mu \left( \frac{D+1}{2}, \frac{1}{2} \right),$$ (1)

where $L$ is hypersphere diameter, $I_\mu(p, q)$ is incomplete Bessel function, and $\mu = 1 - \dfrac{l^2}{L^2}$. The distribution $f(l)$ at $D = 3$ (ball) has the form

$$\frac{12}{L^6} l^2 (L-l)^2 (2L+l),$$ (2)

and at $D = 2$ (disk) is

$$\frac{16}{\pi} \frac{l}{L^2} \left[ \arccos \frac{l}{L} - \frac{l}{L} \sqrt{1 - \frac{l^2}{L^2}} \right].$$ (3)

One can show from (1) that for small values $l$ (where the boundary effect may be neglected) the distribution $f(l)$ has an asymptote

$$f(l) \sim l^{D-1}$$ (4)

(for disk $f(l) \sim l$, and for ball $f(l) \sim l^2$). We can assume that such asymptote will conserve for fractal distributions. In fact, such generalization has been made by Grassberger & Procaccia (1983a, b), who have shown that at small $l$ the correlation integral $C(l)$ behaves as power function

$$C(l) \sim l^\nu.$$ (5)

Here the function $C(l)$ is the integral of distribution $f(l)$ of paired separations between the set objects. Grassberger & Procaccia (1983a) have considered several examples and shown that the values of $\nu$ and $D$ are almost equal. Our further considerations will be based on equality $\nu = D$.

Further we do not use the correlation integral $C(l)$, since the differential law $f(l)$ is more descriptive. The preparation of massive of the paired separations allows deal with concrete structures in spatial distribution of objects. This approach is to be perspective just for small samples of objects, since instead of massive of $3N$ coordinates we use the massive of $N(N-1)/2$ paired separations, which are invariants with respect to choice of coordinate frame. For example, for a sample of 200 points we have about 20,000 separations for statistical treatment instead of 600 coordinates.

In this paper, we describe the results of application of the method of paired separations to current sample of gamma-ray burst sources having measured redshifts $z$.

## CALCULATION OF METRIC DISTANCES

We have applied above method to the sample of 201 long gamma-ray bursts with measured redshifts from the Catalogue the Gamma-Ray Burst Online Index (**http://lyra.berkeley.edu/grbox** – author is D. Perley). In order to find the metric distance $r$ to a gamma-ray burst source, we need to give a cosmological model. A choice of the model is essential, because generally the dependence $r(z)$ is non-linear.

Let us consider the standard $\Lambda$CDM model with cosmological parameters $\Omega_M = 0.28$, $\Omega_\Lambda = 0.72$. The distance to an object with redshift $z$ in co-moving coordinates along the line-of-sight may be calculated using by the following formula (Hogg 2000):

$$r(z) = \frac{c}{H_0} \int_0^z \frac{d\xi}{E(\xi)}, \tag{6}$$

Where the function $E(\xi)$ has the form

$$E(\xi) = \sqrt{\Omega_M (1+\xi)^3 + \Omega_k (1+\xi)^2 + \Omega_\Lambda}, \tag{7}$$

$\Omega_M$, $\Omega_\Lambda$, $\Omega_k$ are the cosmological density parameters for matter, energy, and space curvature, $H_0$ is the Hubble parameter, $c$ is the light speed. In this model, the metric distance is calculated through the redshift by the formula (see, e.g., Vasilyev 2008), following from (6) at $\Omega_k = 0$:

$$r(z) = \frac{c}{H_0} \int_{\frac{1}{1+z}}^1 \frac{dy}{\sqrt{y(\Omega_M + \Omega_\Lambda y^3)}}. \tag{8}$$

In addition to the $\Lambda$CDM model, we consider the "tired light" model. In this model, the formula analogous to (8) has the form (see, e.g., LaViolette 1986):

$$r(z) = \frac{c}{H_0} \ln(1+z), \tag{9}$$

and Euclid space model, in which the dependence analogous to (8) and (9) is

$$r(z) = \frac{c}{H_0} z. \tag{10}$$

All distances are measured in the Hubble radii

$$R_H = \frac{c}{H_0}. \tag{11}$$

The functions $r(z)$ are shown in Fig. 1. We have unexpectedly revealed that the dependences for the first two models are similar, although the nature of redshift in those is different (see below).

Thus, these two models are hardly distinguished in some tests to be used in practical cosmology. Evidently, all cosmological models under consideration have asymptote (10) at small $z$.

Although the functions $r(z)$ for the $\Lambda$CDM model and "tired light" model are similar, these two models are radically different. The "tired light" model is stationary, and one can compute the paired separations in it with accuracy till peculiar motions. The $\Lambda$CDM model has both peculiar motions and cosmological expansion (evolution of scale factor). The global outflow influences on the distribution $f(l)$ – that "blows" it, preserving the structure. When one analyzes large galaxy surveys (e.g., $SDSS$), this problem does not stand so sharply, since these objects have small $z$ (in particular, main galaxies in $SDSS$ have $z < 0.2$).

General question for all cosmological models under study in context of analysis of the large-scale structure of the Universe is the evolution of this structure. In frameworks of this paper, we believe possible to neglect the evolutionary effects, because we consider the only asymptotic behavior of the function $f(l)$ at the separations, which are small in comparison with the size of region under study.

ESTIMATES OF FRACTAL DIMENSION FOR DISTRIBUTION OF GAMMA-RAY BURSTS

The dependence between burst length $T_{90}$ and $z$ is shown in Fig. 2. We can see from this picture that a small population of short bursts ($T_{90} < 2^s$) exists at redshifts $z < 1$. According to current ideas, short and long bursts have different nature (see, e.g., Woosley & Bloom 2006). We will only consider the long gamma-ray bursts with $T_{90} > 2^s$.

Let us calculate metric distance $r$ to each gamma-ray burst source. Integral distribution $F(r)$ is given in Fig. 3. We can see from this picture that the function $F(r)$ does change slower than $r^3$. Note that the function $F(r) \sim r^D$ for fractal distribution at dimension $D$. Two effects influence on the behavior of integral distribution $F(r)$: 1) the non-homogeneities in distribution of gamma-ray burst sources; 2) increasing of sample incompleteness when distance from an observer grows. In order to analyze these effects in details, we will use the above function $f(l)$.

We will construct the distribution $f(l)$ of paired separations between the gamma-ray burst sources for every cosmological model under consideration (see above). These distributions are shown in Figs. 4-6 for two limited redshifts $z = 2, 3$. We do not consider more distant objects, as the number of gamma-ray burst sources with measured $z$ do sharply decrease at $z > 3$ (see distributions of redshifts and distances to sources in Fig. 7). One can see from Figs. 4-6 that the distributions $f(l)$ for observed gamma-ray burst sources at small $l$ behave differently than analogous functions for random distribution. Three realizations of random distribution are shown in Fig. 4. Those do not differ practically from each other and strongly differ from observed data. One can conclude that the revealed non-homogeneities reflect the fractal signs in spatial distribution of the gamma-ray burst sources. Note that for spatial object distribution, which has no fractal signs, e.g. object concentration having smooth density profile (such as the polytropic one with index one), the behavior of function $f(l)$ at small $l$ is similar to random distribution, in particular, it has the same power asymptote. So if one observes a composition of spherical clusters having smooth (non-fractal) object distribution, one finds an estimation $D \approx 3$. The significant difference of observed asymptote from the case $D = 3$ evidences the fractal signs or dominating the structures which differ from spherical ones strongly.

The results of power fitting are shown in Table 1. The given values of correlation dimension $D$ have been computed using the asymptote $f(l) \sim l^{D-1}$. The values $D$ for observed data are significantly less than for random distribution. Those are from 2.2 till 2.5 depending on a chosen cosmological model and limited redshift. So one can say that the spatial distribution of gamma-ray burst sources has the fractal signs at the scales $l \sim (0.2 \div 0.3)L$, at least till $z = 2 \div 3$. At lesser scales, a small sample volume acts, at scales comparable with system size, the boundary effect plays an important role.

## SEARCH FOR STRUCTURES IN DISTRIBUTION OF GAMMA-RAY BURSTS

We have selected 100 pairs of objects with the smallest paired separations in 3D space in framework of one model (as an example we consider the standard $\Lambda$CDM model). When we have analyzed the space distribution of pair components, we have found a volume element having an increasing concentration of objects: it is elongated structure with equatorial coordinates $23^h56^m < \alpha < 0^h49^m$ and $19^o < \delta < 23^o$ (constellations Pisces-Andromeda), $z \in [0.81, 0.94]$ (some information on gamma-ray burst sources from this structure is given in Table 2). A rough estimation of probability, that four objects may simultaneously place in such volume everywhere within volume under consideration at random distribution of gamma-ray burst sources, is equal to $\sim 10^{-4}$. The linear size of this structure is $\sim 0.2 R_H \approx 800$ Mpc, that at least double size of "Great Sloan Wall'' (Gott et al. 2005, Park et al. 2005). Possibly, this non-homogeneity is the largest structure within $z < 1$. Every fiftieth gamma-ray burst source with measured $z$ belongs to this structure. Note that a few other gamma-ray burst sources 981226, 980703, 041006, 040924, 091208B (last object has been recently discovered) join to this basic structure formed by four above objects.

Moreover, we have found several pairs of objects, in which the separations between components are essentially less than the mean distance to the nearest neighbor for random distribution of points. The most interesting case is a discovery of two objects 060927 and 060522 having the redshifts 5.467, 5.11 and coordinates $\alpha_1 = 21^h58^m$, $\delta_1 = 5^o$; $\alpha_2 = 21^h32^m$, $\delta_2 = 3^o$. Also we have found two triplets of objects: $14^h40^m < \alpha < 15^h12^m$, $-12^o < \delta < 0^o$, $z \in [2.4, 2.7]$; $12^h40^m < \alpha < 13^h13^m$, $8^o < \delta < 17^o$, $z \in [3.0, 3.6]$. Finally, the group of four objects with coordinates $16^h40^m < \alpha < 18^h40^m$, $40^o < \delta < 60^o$ (constellations Hercules-Draco), $z \in [0.96, 1.29]$ was revealed. If these concentrations correspond to the large-scale structures, then their typical scales are comparable or even more than the sizes of the "Great Sloan Wall''. One may expect that new revealed gamma-ray bursts with measured $z$ will meet more often in these regions than in other sky domains.

In connection with above results, it is of interest to study the distributions of other objects (e.g., quasars) within these regions. We have carried out such study in the first above region: the distribution of quasar redshifts within this plate is plotted in Fig. 8 (it was constructed using the data of 12$^{th}$ Catalogue of Veron-Cetty & Veron (2006), see its description in Veron-Cetty & Veron (2003)). The distribution of $z$ for quasars is non-monotonous: it reaches a maximum at $z \approx 1.2$, that is slightly more than the redshifts of revealed group of gamma-ray burst sources.

## DISCUSSION

Thus, our study of spatial distribution of the gamma-ray burst sources using the distribution $f(l)$ of paired separations allows to find (although with low resolution) the most prominent spatial structures (with sizes up to Gpc), while such analysis of galaxy distribution (e.g., from SDSS) allows to investigate rather thin structure with better resolution. The fractal signs in distribution of gamma-ray burst sources were discovered (fractal dimension $D = 2.2 \div 2.5$).

Note that now the sample of gamma-ray bursts with measured redshifts increases approximately by 40 objects per year. Our representation of results using the current sample (201 objects) is attended by some risk. Nevertheless, we have decided to make this step taking into account the historical experience. The Hubble law was firstly formulated using only 24 galaxies (Hubble 1929), and the $\Lambda CDM$ model in cosmology has appeared due to statistical study of the sample of 42 supernovae (Perlmutter et al. 1999). Later these conclusions were confirmed using more representative samples.

Note that our conclusions are based on two assumptions:

1) gamma-ray burst sources with measured redshifts form a representative sample in spite of their avoidance of the zone around galactic equator;

2) the asymptotic behavior of distribution of paired separations $f(l) \sim l^{D-1}$ is valid also for fractal sets of points.

Moreover it is important to note that our results derived by distribution of paired separations are invariant with respect to a form of the region under study, because we use an asymptotic behavior of this distribution at separations which are much less than the size of this region. Note that we can "mark" only the largest structures, because the sample of gamma-ray bursts with measured redshifts is rather small. Usually, more fine structures (which sizes are less than a typical mean separation between neighbors) will be omitted during "marking". One can hope that a sample of gamma-ray bursts with measured redshifts will significantly increase in near future, it would be possible to localize the marked structures and to "mark" the structures at smaller scales.

The authors thank very much Yu.N. Gnedin, V.N. Yershov, N.G. Makarenko, A.G. Sergeev, V.V. Tsymbal, A.V. Yushchenko for valuable discussion of results. Also we are grateful to N.G. Makarenko for kindly given references to the papers by Grassberger & Procaccia (1983a, b). One of the authors (V.V. Orlov) thanks the President program for support of leading scientific schools (grant NSh-1323.2008.2).

Table 1: Results of power fitting for observed data and random distribution at redshifts $z < 2$ and $z < 3$.

| Model | Observed data | Random distribution |
|---|---|---|
| | $z < 2$, $N = 119$ | |
| Euclid space | D = 2.18 ± 0.07 | D = 2.92 ± 0.32 |
| "Tired light" | D = 2.46 ± 0.10 | D = 2.87 ± 0.21 |
| ΛCDM model | D = 2.37 ± 0.09 | D = 3.06 ± 0.25 |
| | $z < 3$, $N = 160$ | |
| Euclid space | D = 2.21 ± 0.08 | D = 2.94 ± 0.39 |
| "Tired light" | D = 2.50 ± 0.12 | D = 3.04 ± 0.23 |
| ΛCDM model | D = 2.43 ± 0.11 | D = 3.04 ± 0.20 |

Table 2: The group of gamma-ray burst sources, which has been found by method of near pairs.

| Numbers of gamma-ray bursts | $\alpha$ | $\delta$ | $z$ |
|---|---|---|---|
| 080710 | $0^h33^m$ | +19°30' | 0.845 |
| 060912A | $0^h21^m$ | +20°58' | 0.937 |
| 051022 | $23^h56^m$ | +19°36' | 0.809 |
| 050824 | $0^h49^m$ | +22°37' | 0.83 |

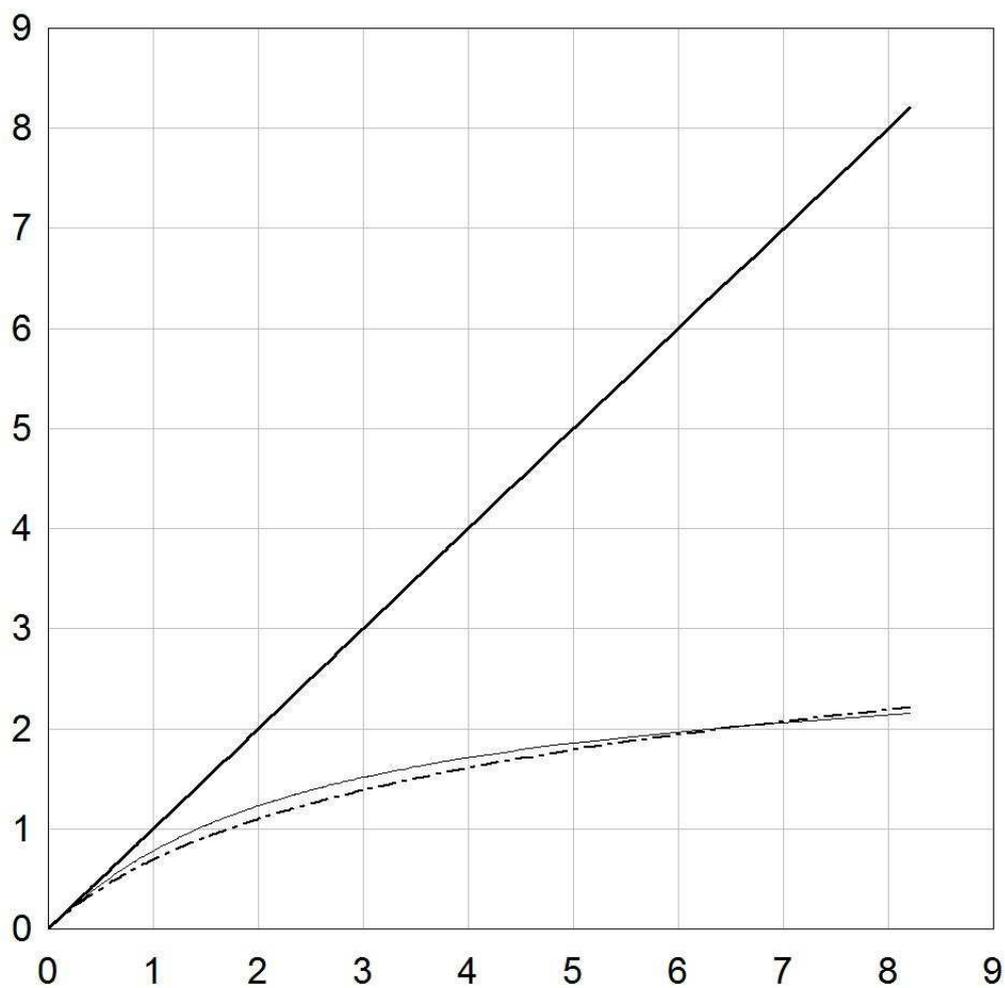

Figure 1: Dependencies of object metric distance of redshift for three cosmological models: solid thick line corresponds to the Euclid space, solid thin line – to the $\Lambda CDM$ −model, dot-dashed line – to the "tired light'' model.

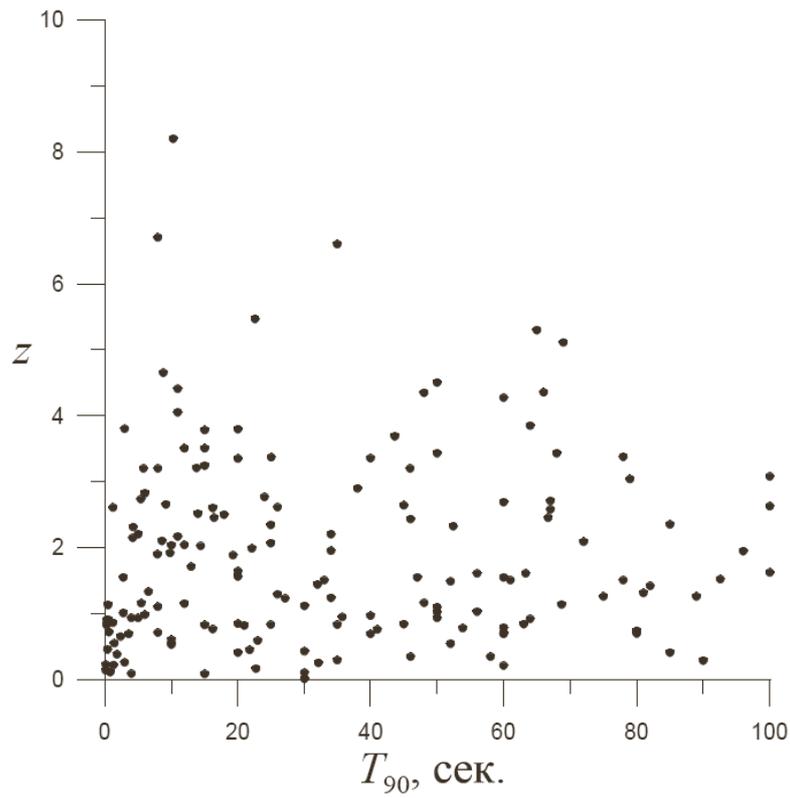

Figure 2: Dependence of the burst length $T_{90}$ on $z$.

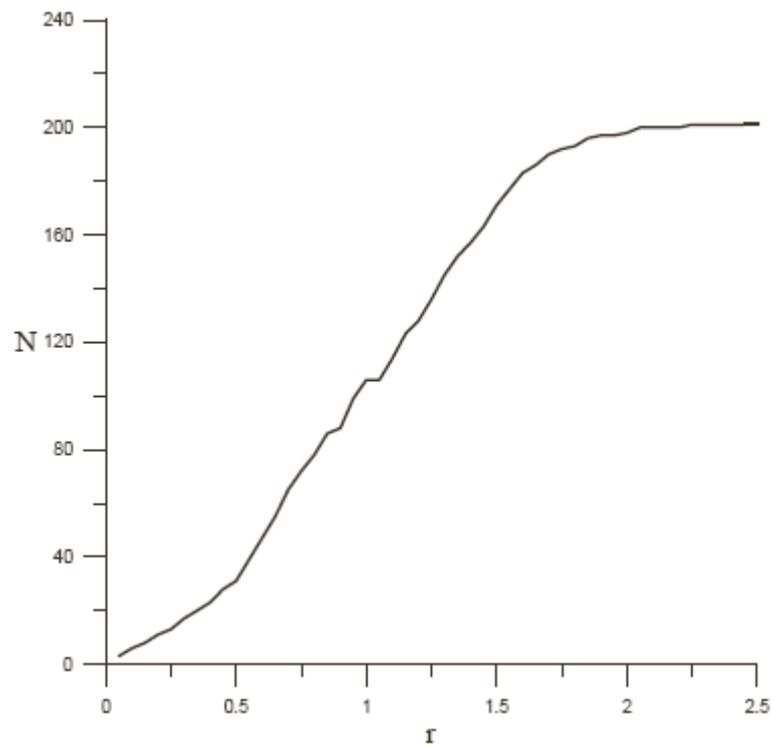

Figure 3: Integral distribution of the gamma-ray bursts by the metric distance in the "tired light" model.

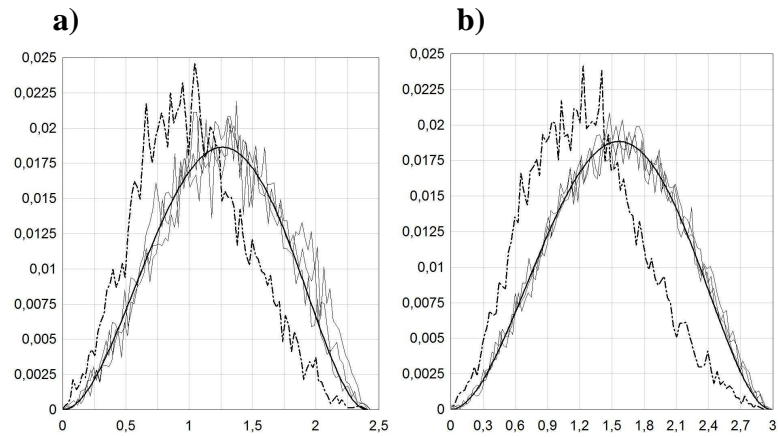

Figure 4: Distributions $f(l)$ of paired separations for the gamma-ray burst sources with measured $z$ (ΛCDM model): a) $z < 2$ ($N = 119$); b) $z < 3$ ($N = 160$). Also three different realizations of distribution $f(l)$ for random distribution of $N$ points inside the ball are drawn by thin solid lines. The solid thick line corresponds to analytical function $f(l)$ for Poisson distribution.

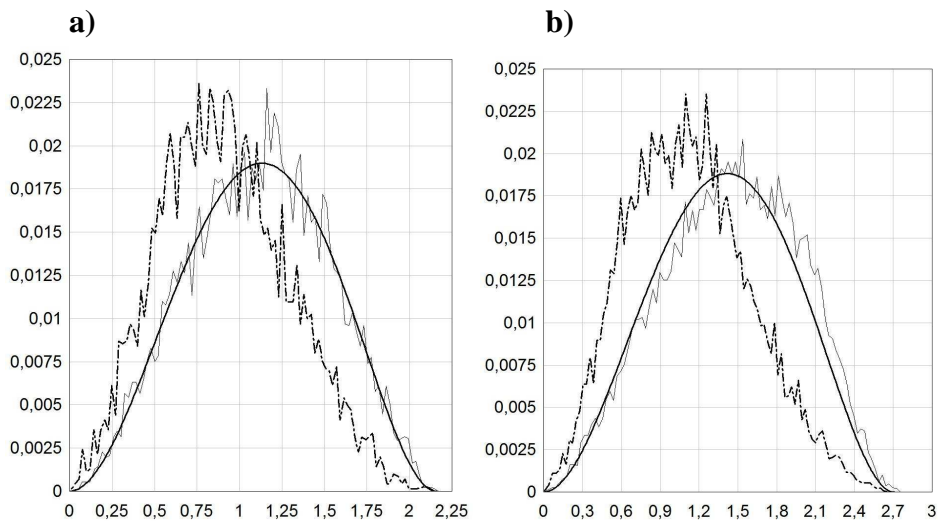

Figure 5: The distributions $f(l)$ of paired separations for the gamma-ray burst sources with measured $z$ ("tired light" model): a) $z < 2$ ($N = 119$); b) $z < 3$ ($N = 160$). Also the distribution $f(l)$ for random distribution of $N$ points inside the ball are drawn by thin solid lines. The solid thick line corresponds to analytical function $f(l)$ for Poisson distribution.

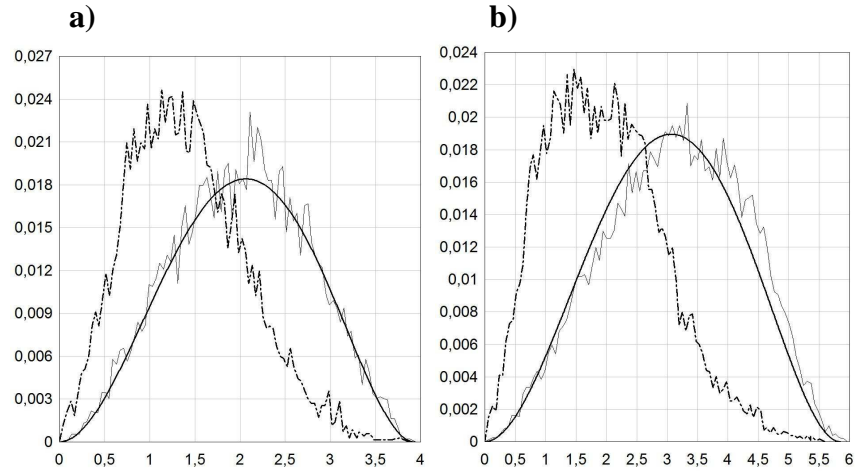

Figure 6: The distributions $f(l)$ of paired separations for the gamma-ray burst sources with measured $z$ (Euclid space): a) $z < 2$ ($N = 119$); b) $z < 3$ ($N = 160$). Also the distribution $f(l)$ for random distribution of $N$ points inside the ball are drawn by thin solid lines. The solid thick line corresponds to analytical function $f(l)$ for Poisson distribution.

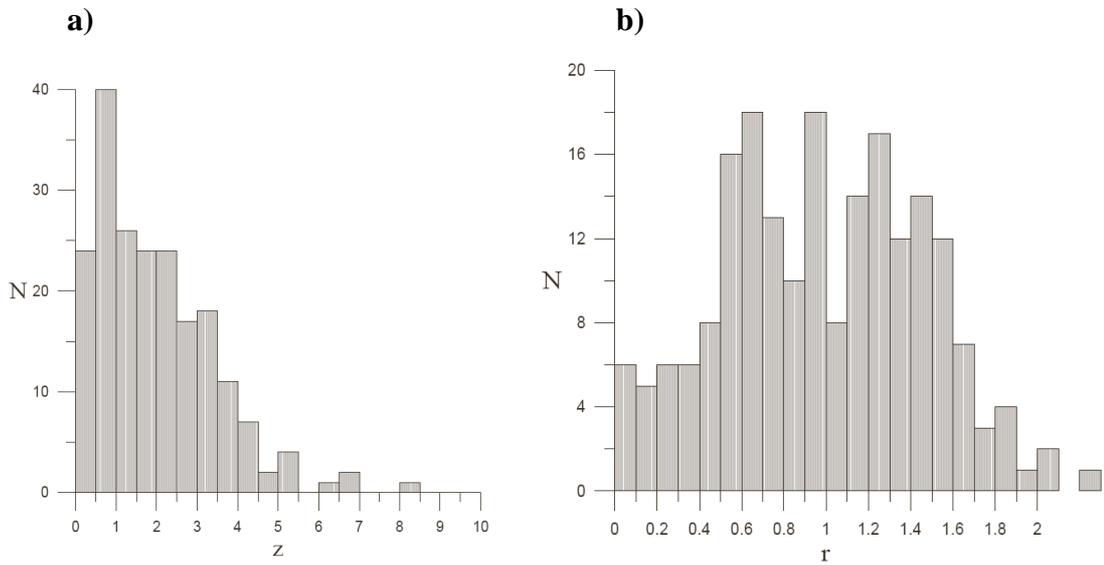

Figure 7: The redshift distribution of the gamma-ray burst sources (a) and their distances in the "tired light" model (b).

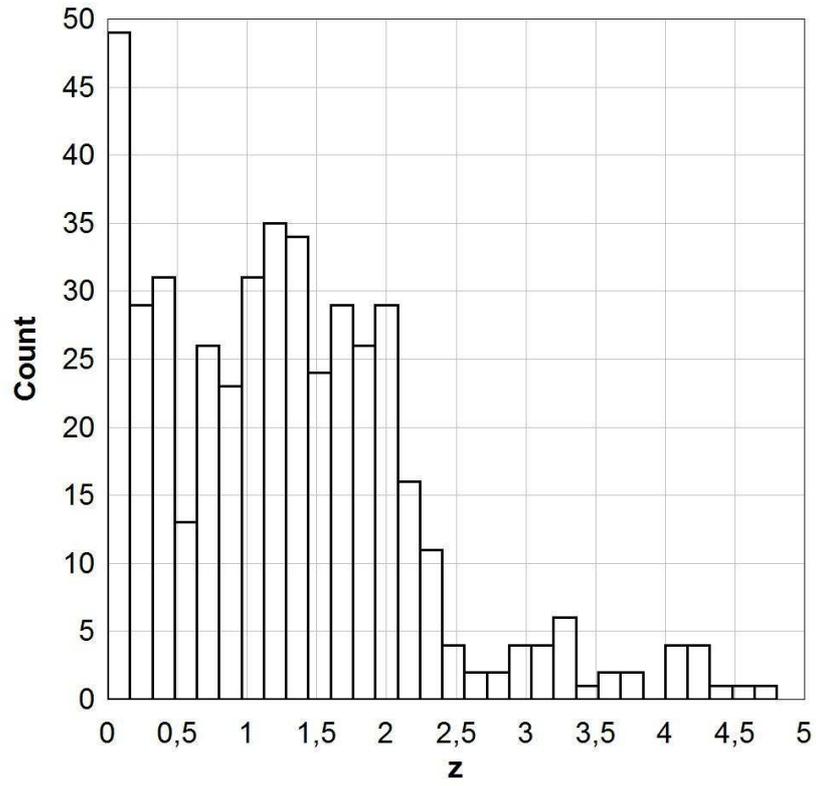

Figure 8: The redshift distribution for the quasars within the plate corresponding to the high concentration of the gamma-ray bursts sources.